\documentclass{article}

\usepackage{amsmath}
\usepackage{amssymb}
\usepackage{hyperref}

\title{On the Heritage of Crypto Assets --\\
       Tales From the Crypt Protocol}

\author{Fr\'ed\'eric Prost}
\date{%
       frederic.prost@univ-grenoble-alpes.fr \\
       LIG - Univ. Grenoble Alpes, France\\
       \today}

\begin{document}

\maketitle

\begin{abstract}
  We discuss some issues to the inheritance of crypto assets. We propose a distributed, privacy preserving, protocol to establish a consensus on the death of the owner of crypto assets: the Tales From the Crypt Protocol. Until the actual death of the owner no link can be made between public information and the corresponding crypto assets. This protocol is generic and could be incorparated into any arbitrary crypto platform.  
\end{abstract}

\section{Introduction}
  \label{sec:introduction}
 
   Cryptocurrencies, starting with Bitcoin \cite{Nakamoto_bitcoin}, have shaken the world of finance in less than a decade. Moving from a pipe dream idea to an every day reality in the meantime. At the time of writing the total market capitalisation of BTC is around 10\% of the market capitalisation of gold. There are many discussions on the nature of money, and assessing the relative merits of BTC vs gold as a store of value. One of the central feature for any store of value is seldom discussed though: the issue of inheritance. This is a proposal to discuss and address some technical problems linked to inheritance.

   Among the mandatory properties that a store of value must have, the heritability property is a central one. On a long enough timeline the survival rate for everyone drops to zero. Transmitting wealth to the next generations is not a peripheral issue, nor one that you can dodge. The body of laws, stories and traditions about inheritance is immense. In fundamental texts like the Bible \cite{Bible} or the Odyssey \cite{Odyssey}, the question of who inherits what from whom, and more generally all kinds of problems linked with succession, are major preoccupations. 

  The issue of inheritance is orthogonal to the actual implementation of the store of value. Society, in a very broad sense, is the tool traditionally used to transfer titles and to settle questions like: "Who is the new king?". Regarding material wealth, objects do not disapear when you die. These remarks no longer hold with cryptocurrencies. Indeed, one fundamental feature of crypto assets is that no one but the owner of the appropriate keys can transfer wealth. The ownership of crypto assets amounts to the knowledge of the keys and vice-versa: everyone that knows the keys is deemed to be the rightful owner of the associated assets. But by definition, and under these circumstances, one cannot actually implement his/her own succession because the knowledge of the keys disapear with their death. It appears that we are finding ourselves painted into a corner: a good crypto asset platform can only allow transfers initiated by the legitimate owner; and the owner cannot transfer anything once dead, making the succession of crypto assets seemingly impossible.  

  In this paper we discuss various issues linked to the inheritance of crypto assets. We propose a solution to the specific issue of acknowledging the death by the network. This acknowledgment can later be used as a trigger for succession transactions and contracts. This is the "Tales From the Crypt Protocol". This protocol respects privacy and is distributed, in the original spirit of cryptocurrencies. Many unresolved issues remain to be solved to cope with all the intricacies linked to the practice of heritage. We hope that this work will open a fruitful research activity and will inspire others to progess on this subject.      
   
\section{The issues of the inheritance of cryptocurrencies}
   \label{sec:inheritance_crypto}

   \subsection{Inheritance and crypto assets, a short review of problems to be solved}
   \label{subsec:problem_review}

   Let's examine some issues, as well as some workarounds, raised by the issue of crypto assets inheritance. A one liner frame of the question to be addressed goes something like this: "How can my seven years old daughter inherits the content of my crypto wallet?". This is a starting point, there are many subtler subproblems. Actually the general problem of inheritance becomes more and more complex the more you consider it seriously. Let's examine a sample of those issues, together with some tentative solutions, by increased level of complexity. 

   \begin{enumerate}
        \item The first idea to solve the basic ''seven years old daughter inheritance'' problem is: (a) - to set up a meeting with a lawyer. (b) - To write down the wills on a document, including the appropriate private keys. (c) - To seal off the enveloppe.  (d) - To hope for the best. 

 	      This natural solution presents many challenges. The more salient being that Bitcoin has been built precisely to provide trustless agreements. There is maybe nothing as opposed to this aim than having to go to see a lawyer, and having to rely on the professional integrity and competence of this lawyer. This is a poster child of all the issues linked with centralization. I am not even touching on the additional issues of anonymity, risks (for lawyers that will be targeted by wrongdoers if this practice become mainstream, for your wallets...) etc. The saying "not your keys not your coins" sums it all. Essentially this solution reintroduces the single point of failure.   

	\item The second idea that may come to mind is to put all the keys on a thumbdrive, or write them down on a piece of paper, and lock them into a safe at home. It marginally improves on the previous point if you have more trust in your family. Besides the hazards that such a practice would produce if it were widely adopted, it has the following drawback: Actually my seven years old daughter is not my only heir. Let's say I have four kids and seven nephews between whom the inheritance is to be divided. It is not as if dramas about succession, struggles within families, and communities, are a literary genre in their own right. Moreover, how can you be sure that the one opening the safe will behave correctly? It is harder to cheat with a pile of physical gold because there may be witnesses, the material has to be moved etc. With crypto assets you just have to remember a passphrase. No one can stop you from using it later. No one can delete this passphrase from your memory. 

	\item It is possible to be smarter and to write a smart contract that implements the succession wills. It solves the "four kids, seven nephews" problem. But it raises a new problem: how will the blockchain be aware of the death of the owner of the smart contract? This is a variant on the famous oracle problem \cite{Oracle}. Moreover, the heirs may not be of age to understand the technology, nor to have the legal rights to access such kind of funds. Some of the heirs may also not have wallets in the first place. If so the mechanisms by which the proper credentials could be transmitted to them, without being compromised, remain mysterious. 

       \item There is another issue: what if the four kids and the seven nephews die with the one they are supposed to inherit from? Let's say, for instance, that they all disappear simultaneously in a plane crash. It is not possible to re-write the smart contract. The heritage disappears (more precisely it becomes inaccessible) in such a scenario. In real life there are specific laws and legal practices to deal with such kind of situation. 

       \item A rather simple solution is to set up an equivalent of a time capsule. If a date is chosen sufficiently far enough in the future, then the death of the capsule owner becomes a certain event. It can be done via smart contracts that just have to wait until some block number is reached in the blockchain before being executed. The drawbacks lie in the lack of flexibility and the necessary approximation of the time of death. A middle-aged victim of a traffic accident could potentially lead to a succession process stalled for more than half a century. Moreover, the probability that the potential beneficiaries of the inheritance may have died too in the meantime increases. 

       \item An improvement over the previous idea is to use a dead man's switch. Instead of using the maximum age plus a safety margin for the time capsule deadline, it is possible to use a shorter frame. If necessary one has just  to edit the time capsule deadline before it is executed. Of course the death of the time capsule owner stops this process of reprogramming, and the time capsule is eventually delivered. It, partially, solves the issue of the lag between death and succession. On the other hand it requires a constant vigilance and work. 

        \item Everyone is going to die but we hope it will be as last as possible. Life execptancy has improved a lot lately. From a practical point of view it is a very challenging aspect of succession to manage. It is very difficult to anticipate the technological environment in a few decades. However, a credible proposal for succession must be resistant to the future. It suggests that any solution should be integrated within the  crypto platform itself rather than relying on outsourced processes.    
   \end{enumerate} 

   \subsection{Existing solutions review}
   \label{subsec:existing_solutions}

\begin{itemize}
   \item Sarcophagus \cite{sarcophagus} is a dead man switch implementation that is blockchain-enabled. It is resistant to censorship and immutable. It is done by the combination of Arweave \cite{Williams2017ArchainAO} for a permanent storage of data, and Ethereum to support the ERC20 Sarco Token. This token is used to pay so called archeologists which are in charge of releasing the data (essentially an encrypted file) to the person of interest. The user have to select one or more existing archeologists. The archeologist public key is used as an outer layer of encryption. This outer layer has to be rewrapped at predefined dates in the future. If one date expires then the archeologist decrypts the outer layer. The inner layer is the data encrypted with the public key of the final receiver that can decrypt it.    

   \item Ternoa \cite{Ternoa} is a french start-up that proposes a "death protocol" which is basically a smart contract triggered by the API's of local authorities registering deaths. It presents the problem of being a centralised solution. One issue is that it is easier to hack the local authorities database (or to bribe agents working for this agency) than to break a distributed solution relying on crypto technologies. Another issue is that there is no standard API to deal with this issue that is shared amongst countries. Each solution is limited to one nation-state at best. Finally there is no warantee that the API are not going to change in the future.   

    \item Casa \cite{Casa} is a company that proposes solutions based on mutli-signature schemes. Their primary service is to provide better resiliency for crypto wallets. They also have an inheritance product that is basically a technological implementation of the second bullet point examined in section  \ref{subsec:problem_review}. 
\end{itemize}

\section{The Tales From the Crypt Protocol}
   \label{sec:tfcp}
 
 The Tales From the Crypt Protocol (TFCP in the rest of this document) is a distributed, privacy preserving, uncensorable, open protocol designed to produce a consensus mechanism linked to the death of a physical person. The TFCP could be incorporated into arbitrary blockchains, modulo the governance peculiarities of the considered blockchains. 

 We propose to introduce new kind of transactions that follow specific rules. This proposition is justified by the following motivations and interests:
\begin{itemize}
   \item Everyone is going to die eventually. Therefore, it makes sense to consider a special case for such an event. Maybe you are never going to make crypto transactions, but what is clear is that one day you are going to die.
   \item The death happens only once. 
   \item There is no universally standardized service or norms for death registration. Every country has its own administrative processes.
   \item It is relatively easy to hack hospitals or morgues IT systems. There are many entry points, and many levels at which the system can be compromised. To have a truly decentralized mechanism to acknowledge the death makes the system more resilient to fraud. The aim is to build a consensus mechanism for this specific event akin the distributed consensus on a public ledger.  
   \item Inheritance transactions fundamentally differ from usual transactions because, by definition, they cannot be performed by the owner of the account since they are performed after his/her death. In most cases this event cannot be forecasted precisely.
   \item Adding a special case for inheritance transactions opens the possibility to make existing cryptocurrencies able to evolve and integrate them. How this integration can take place depends on the specifics of the governance of the considered crypto platform.  
\end{itemize}   

   \subsection{TFCP scheme}
   \label{subsec:tales_scheme}

   In this section we describe the TFCP without going into technical details. We focus on the general ideas and rationale behind the protocol. The TFCP involves two sets of actors, {\bf Registrars} and {\bf Witnesses}. This is a fundamental mechanism to provide privacy properties. The TFCP also relies on adversarial incentives (rewards and penalties)  to prevent bad actors from interfering with the desired behavior of the protocol. Heritage details are not considered in TFCP. The sole purpose of the TFCP is to provide a signal that is equivalent to the recognition of the death of the {\bf Donor} by the network. We say that the {\bf Acknowledgment} has been enacted by the network.

   \subsubsection{Definition of terms and concepts used in the TFCP}
   \label{subsubsec:terms}
   Let's start by introducing the actors considered. Here we identify persons with the secret key of an account, and with the account itself. We are using the term "account" to denote both the physical person, the private keys, and the associated wallet. When necessary, the distinction between the physical person, the private keys and an the associated wallet is explicitly stated. 
   \begin{itemize}
	\item The {\bf Donor} is the account of the physical person for which the network has to enact the death {\bf Acknowledgment}.   
        \item The {\bf Security Deposit} is the account used to signal the death of the {\bf Donor}. The {\em Donor} has the key of this account.
	\item The {\bf Witnesses} is the set of accounts that testify on the death of the {\em Donor}. 
        \item The {\bf Registrars} is the set of accounts that provide public keys and are operative to share secrets. 
   \end{itemize}

Let's follow by the definition of values and terms that play a special role in TFCP. 
\begin{itemize}
       \item The {\bf Threshold} is the amount of coins that has to be stacked on the {\bf Security Deposit} account by the {\bf Witnesses} to signal the death of the {\bf Donor}.   
       \item  The {\bf Wills} is a document that contains the link between the {\bf Donor}, the {\bf Announcement} and the {\bf Security Deposit}. Those links are not public, they are encrypted using a shared key between {\bf Registrars}. It is signed by the {\bf Security Deposit}. 
       \item  The {\bf pre-Wills} is very similar to the {\bf Wills} document. Both can be considered equivalent at this level of abstraction. {\bf pre-Wills} are published to recruit {\bf Registrars}. Once the recruitment, and some checks, are done {\bf pre-Wills} are transformed into {\bf Wills} (mainly by stripping away technical informations), and published. Both the {\bf pre-Wills} and the {\bf Wills} are signed by the {\bf Security Deposit}. 
         \item The {\bf Announcement} is a public document that contains the {\emph social security name} of the {\bf Donor}, the {\bf Fees} to be distributed to the {\bf Witnesses}, the {\bf Threshold}, the {\bf Deliberation Time} and the address of the {\bf Security Deposit} account.
        \item The {\bf Ante} is the amount of coins that a {\bf Witness} has transferred to the {\bf Security Deposit} to signal the death of a {\bf Donor}.  
        \item The {\bf Shares} are the share of a secret share scheme that is used by the {\bf Donor} to multi-encrypt the {\bf Wills}.
	\item The {\bf Fees} are the rewards for the participation of both the {\bf Registrars} (specified in the {\bf Wills}) and the {\bf Witnesses} (specified in the {\bf Announcement}). 
	\item The {\bf Bail} is the amount of coins that {\bf Registrars} have to stake on a special account. They are there to insure that {\bf Registrars} are executing the protocol in a fair way.
	\item The {\bf Deliberation Time} is the amount of time during which the {\bf Wills} account remains locked after the {\bf Threshold} has been reached. 
        \item The {\bf Acknowledgment} is the reckoning of the death of the {\bf Donor} by the network. 
\end{itemize}

The schematical use case, if everything goes as planned, unfolds as follows: 
\begin{enumerate}
   \item The {\bf Donor} selects a set of {\bf Registrars}.
   \item The {\bf Donor} publishes the {\bf pre-Wills} under the identity of the {\bf Security Deposit}.  
   \item The interested {\bf Registrars} check the validity of the published documents. They can accept the {\bf Donor}'s request by publishing their acceptance or decline (by doing nothing). 
   \item Once enough {\bf Registrars} have accepted, the {\bf Donor} sends a {\bf Share} of a secret key to each interested {\bf Registrar}. 
   \item The {\bf Donor} encrypts  the {\bf Wills} with the shared key of step (4). The {\bf Wills} are published under the {\bf Security Deposit} identity. 
   \item The {\bf Donor} publishes the {\bf Announcement} under the {\bf Security Deposit} identity. 
   \item When the {\bf Donor} dies, the {\bf Witnesses} transfer coins to the {\bf Security Deposit} account. 
   \item When the {\bf Threshold} is reached on the {\bf Security Deposit} account, the process of acting the death {\bf Acknowlegdment} by the network is initiated. 
       
         Two cases: either there is a move from the {\bf Donor}'s account before the {\bf Deliberation Time} has elapsed or not. 
  
         \begin{enumerate}
            \item If there is no move.
             \begin{enumerate}
               \item The {\bf Registrars} decrypt the {\bf Wills}. A public version of the {\bf Wills} is published by the {\bf Registrars}.  
               \item The {\bf Donor}'s death {\bf Acknowlegement} is acted by the network. 
               \item The {\bf Fees} are transferred to the {\bf Registrars} and {\bf Witnesses} following what has been specified  both in the {\bf Announcement} and the {\bf Wills}. Then, the heritage transfers are done (typically via the execution of smartcontracts), details of which are not in the scope of this paper. 
\end{enumerate}  
            \item If there is a move. Then, the {\bf Donor} is alive and the {\bf Acknowledgment} cannot be acted by the network. It becomes possible to make moves from the {\bf Security deposit}. The {\bf Donor} may initiate a new instance of TFCP.  
         \end{enumerate}
\end{enumerate}

Some remarks regarding this protocol:
\begin{itemize}
    \item The {\bf pre-Wills} contain enough information so that only the {\bf Donor} can produce the {\bf Wills}. It is a bit tricky because at the publishing time of the {\bf pre-Wills} it is not possible to check this link. It becomes possible once enough {\bf Registrars} have accepted and were given a share of the secret key. 
    \item There is nothing preventing the {\bf Donor} to set up as many instances of TFCP's as desired. Since any TFCP has  public partis, the {\bf Announcement} and the {\bf Wills}, it is easy to find the most recent one. It will be the only one considered valid. It makes possible to have several versions of the {\bf Wills}. It implies that a fixed fraction of the {\bf Fees} for the {\bf Registrars} has to be payed immediately, in order to deter useless TFCP instances and unpaid work for the {\bf Registrars} (since all but one TFCP is going to be completed). It could be done via the {\bf Security Deposit} account of the appropriate TFCP instance.  
\end{itemize}

   \subsection{Expected Properties of TFCP}
   \label{subsec:properties}

The TFCP does not solve all the issues discussed in section \ref{sec:inheritance_crypto}. Though, it is expected to have the following good properties:
\begin{enumerate}
   \item Distributed Oracle: the only requirement for {\bf Witnesses} is the amount of {\bf Ante} they are going to transfer to the {\bf Security Deposit}. It ensures a maximally distributed system. There is no special person or organisation. It is  a completely open process. Likewise there are no special requirements for {\bf Registrars}, apart from the {\bf Bail} they have to provide and stack. It is essentially the same kind of requirements than the ones required for the {\bf Witnesses}. Though it is more significant in volume, and unlike {\bf Witnesses} there is less/no freedom in the volume chosen. It is also supposedly for longer time period. {\bf Witnesses} are refunded after the completion of the TFCP instance they are participating in while {\bf Registrars} are not refunded immediately.   
   \item Anonymity: If all actors are honnest but curious they can't link the {\bf Security Deposit} to the corresponding {\bf Donor}'s account before the {\bf Acknowledgment} is enacted by the network: the {\bf pre-Wills} doesn't containe enough information. {\bf Registrars} can make this link if they cooperate after the publication of the {\bf Wills}. It is possible to add an extra layer to build an equivalent of a mix network \cite{Chaum81} to jam this track. Basically {\bf Registrars} can act as mixes in a mix-network so that it is not possible to directly trace back the path between the {\bf Donor} and the {\bf Announcement}. Something like the Monero can also be considered.  Such a kind of solutions introduce costs and complexities at many levels: trade-offs have to be considered on a case by case basis. 
   \item Blockchain agnostism: in principle the TFCP can be adapted to any crypto platform. As discussed at the beginning of section \ref{sec:tfcp}, death is an exception that eventually occurs exactly once for every human. It is not unreasonable to integrate such kind of exceptions in every crypto platform. 
\end{enumerate}

   \subsection{Discussion on the incentive structure}
   \label{subsec:analysis_incentives}

The basic incentive structure of TFCP relies on the following pair of adversarial incentives:
\begin{itemize}
   \item Positive incentives: the {\bf Fees} for {\bf Registrars} and {\bf Witnesses}. 
   \item Negative incentives: the {\bf Ante} and the {\bf Bails} respectively stacked by the {\bf Witnesses} and the {\bf Registrars}. The difference between the two being that the {\bf Ante} are a one-time thing. Moreover, the time horizon for the staking of the {\bf Ante} is much shorter than the one for the {\bf Bails}. Typically the {\bf Deliberation Time} gives an idea of how long the {\bf Ante} will be freezed on the {\bf Security Deposit}. The duration of the {\bf Bails} could be chosen by the market: when the {\bf Donor} is chosing a set of {\bf Registrars} he can choose the {\bf Registrars} with a sufficiently remote published date on their {\bf Bails}.  
\end{itemize}

The delay and precise circumstances to adjudicate the enforcement of the negative incentives have to be tested in order to discover the best trade-offs. Here are some ideas related to these issues:

\begin{itemize}
   \item Alice could create an essentially empty wallet, be minimally active, and transmit the secret key this shallow account to Bob. Then Alice can behave as a {\bf Donor} and set up a TFCP instance. When Alice dies, Bob can make a move on Alice's {\bf Security Deposit} account. The TFCP protocol is halted and Bob can take control of the {\bf Security Deposit} account. Notice that Alice has to actually die for this attack to work because no one can force any {\bf Witnesses} to transfer any {\bf Ante} to the {\bf Security Deposit}. 

         As the name of the {\bf Donor} will be made public it will tarnish his reputation. Another factor that may limit this type of attack is to consider the {\bf Ante} as a bet on honnesty: the appropriate odds will be ultimately set by the market. A null {\bf Deliberation Time} would render this attack impossible, but at the cost of losing the adjudication period.
     
   \item A very wealthy {\bf Witness}, or a flash loan attacker \cite{qin2021attacking}, could, all alone, trigger the {\bf Heritage}  by transfering an {\bf Ante} as large as the {\bf Threshold} to the {\bf Security Deposit}. The attacker would like to obtain the {\bf Fees}. One way to circumvent this attack is to require that $x>1$ separate transactions have to be done over a sufficiently long span of time (otherwise the wealthy attacker could set up mutliple accounts) in order for the TFCP to be valid. Typically something to the order of the {\bf Deliberation Time}. Having a non 0 {\bf Deliberation Time} introduces the risk (from the attacker point of view) of losing the {\bf Ante}. One circumstance where it could be dangerous is if the {\bf Donor} cannot access to his/her account (because of illness or whatever peculiar situation) during the {\bf Deliberation Time}. 
\end{itemize}

\section{TFCP technical specifications}
   \label{sec:tales_technical}

   As a first approximation all public documents are published on the blockchain (or a commitment mechanism has to be set up). Every public document is signed with the keys of the publishing account. This account pays the fees of publication.  

   \subsection{Registrars}

   To be registered as a {\bf Registrar}, an account has to commit itself to stack a given amount of coin on a {\bf Bail} account. Moreover, the {\bf Registrar} has to declare for how long the {\bf Bail} is due. Precise amount and minimal bailing time are determined through the governance of the considered blockchain.    

   Typically {\bf  Registrars} will be institutions since their life expectancy can be longer than human's one. There is a kind of chicken and egg situation: {\bf Registrars} may be forced to bequeath their portfolio. This is a special case of inheritance (because {\bf Registrars} are not required to die), and each crypto platform have to adopt the suitable policies to implement solution to this issue. 

   \subsection{Wills}

   The {\bf pre-Wills} are published by the {\bf Security Deposit} and contain at least:
   \begin{itemize}
	\item The {\bf Donor}'s account address encrypted using a shared secret $s$ a key.  
        \item The {\bf Security Deposit}'s account address. 
	\item The number $n$ of {\bf Shares} needed to compute the shared key $s$.
        \item A list $\{R_i | 1 \leq i \leq m\}$, with $n > m$, of acceptable {\bf Registrars}. This is optional. If there is no list specified, then any {\bf Registrar} can apply to certify the {\bf Wills}.
        \item The ${\bf Fee}$ for the {\bf Registrars}. 
	\item The {\bf Donor}'s signature of the hash of the {\bf Wills}.
   \end{itemize}

Notice that the {\bf Donor}'s signature cannot be checked by the {\bf Registrars} when they receive the {\bf pre-Wills}. Only the signature of the {\bf pre-Wills} by the {\bf Security Deposit} account can be checked. The validity of the {\bf Donor}'s signature can be checked once the {\bf Acknowledgment} has been acted by the network. 

The difference between the {\bf pre-Wills} and the {\bf Wills} is the that the list of acceptable {\bf Registrars} is not part of the {\bf Wills}. 

   \subsection{Annoucement}

    The {\bf Announcement} is a public document. It is signed by the {\bf Security Deposit}. The {\bf Annoucement} contains~: 
   \begin{itemize}
        \item The civil name of the {\bf Donnee} and enough additional information to be identified in the real life. It may include, but is not restricted to: birth date, birth place, middle names, social security number etc. 
        \item The reference to the {\bf Wills}. Typically the bloc number on which the {\bf Wills} have been published.
	\item The {\bf Security Deposit} address on the blockchain.
	\item The {\bf Fees} for the {\bf Witnesses}.
	\item The {\bf Deliberation Time}.  
   \end{itemize}

\section{Conclusion}
   \label{sec:conclusion}
   In this paper are presented some ideas to tackle with the issue of inheritance. It is focussed on a distributed way to initiate the heritage process. The TFCP is agnostic with relation to the crypto platform considered. How to integrate it and technical details are issues of governance of interest. THCP introduces special kind of transactions and procedures. THCP doesn't require any kind off rollback, the {\bf Security Deposit} is playing the role of a backup account if there were problems (misinformation or hacking attempts). The TFCP respects privacy until the death of the {\bf Donor} . Unless {\bf Witnesses} conspire there is no way to link the publicly known information, ie the name of a {\bf Donor}, to the actual {\bf Donor} account. This link will appear after the death has been acknowledged by the network. It is possible, using a Mix-net kind of idea \cite{Chaum81}, to reach some type of anonymity at the expense of simplicity.

   \subsection{Open questions}
   \label{subsec:open_questions}
   There remains many unsolved challenges with relation to the inheritance of crypto assets. We list a few of them hoping that they will be tackled by some of the readers. 

   \begin{itemize}
        \item Can {\bf Registrars}  resell/transfer their accepted tasks? Hopefully the inheritance is set to be triggered a long time ahead. What happens if {\bf Registrars} have died (filed for bankrupcy) in the meantime ?
        \item  Shall a market for {\bf Registrars} actions be made? For instance we could decide that the first ones to decrypt the {\bf Wills} get more rewards than the remaining ones in order to accelerate the process.  
	\item Transferring credentials to the next generations: heirs can be too young to understand how crypto assets work. What happens before they come of age? How are the credentials stored and safely delivered when majority (or the suitable time) has come? Many real life scenarios have to deal with such considerations.  
	\item The heirs are not precisely known in advance. Think at the holiday travel accident scenario during which a whole family is victim for instance. Is it possible to delay the triggering of the heritage until some form of resolution has been settled? 
   \end{itemize} 

  The right tuning of the incentive structure (importance of the {\bf Fees}, average acceptable {\bf Deliberation Time} etc.) is going to be found via market mechanisms and progressively discovered through use. It can hardly be anticipated before being tested in real life and at large scale.

\bibliographystyle{plain}
\bibliography{biblio}

\end{document}